\documentclass[12pt,journal,letterpaper,draftcls,compsoc,onecolumn]{IEEEtran}

\usepackage{multirow}
\usepackage{amssymb}
\usepackage{graphicx}
\usepackage{amsmath}
\usepackage[noadjust]{cite}
\usepackage[tight,footnotesize]{subfigure}
\ifCLASSINFOpdf
\else
\fi
\begin{document}
%
\title{An Optimal Game Theoretical Framework for Mobility Aware Routing in Mobile Ad hoc Networks}
%
%
%

\author{Mehrdad~Khaledi,
        Mojgan~Khaledi,
        and~Hamid~R.~Rabiee,~\IEEEmembership{Senior Member,~IEEE}
\IEEEcompsocitemizethanks {\IEEEcompsocthanksitem Mehrdad Khaledi is with Computer and Systems Engineering Department, Rensselaer Polytechnic Institute. E-mail: khalem@rpi.edu.
\IEEEcompsocthanksitem Mojgan Khaledi is with School of Computing, University of Utah. E-mail:mojgankh@cs.utah.edu.
\IEEEcompsocthanksitem Hamid R. Rabiee is with Department of Computer Engineering, Sharif University of Technology, Tehran, Iran. E-mail: rabiee@sharif.edu}
\thanks{}}
\IEEEcompsoctitleabstractindextext{
\begin{abstract}
Selfish behaviors are common in self-organized Mobile Ad hoc Networks (MANETs) where nodes belong to different authorities. Since cooperation of nodes is essential for routing protocols, various methods have been proposed to stimulate cooperation among selfish nodes. In order to provide sufficient incentives, most of these methods pay nodes a premium over their actual costs of participation. However, they lead to considerably large overpayments. Moreover, existing methods ignore mobility of nodes, for simplicity. However, owing to the mobile nature of MANETs, this assumption seems unrealistic. In this paper, we propose an optimal game theoretical framework to ensure the proper cooperation in mobility aware routing for MANETs. The proposed method is based on the multi-dimensional optimal auctions which allows us to consider path durations, in addition to the route costs. Path duration is a metric that best reflects changes in topology caused by mobility of nodes and, it is widely used in mobility aware routing protocols. Furthermore, the proposed mechanism is optimal in that it minimizes the total expected payments. We provide theoretical analysis to support our claims. In addition, simulation results show significant improvements in terms of payments compared to the most popular existing methods.
\end{abstract}

}
\maketitle
\IEEEdisplaynotcompsoctitleabstractindextext
\IEEEpeerreviewmaketitle
\section{Introduction}
%
%
%
%
\IEEEPARstart{M}{obile} Ad hoc Networks (MANETs) are dynamically reconfigurable wireless networks in which mobile nodes communicate with each other without the need for any fixed infrastructure. In recent years, with proliferation of mobile communicating devices, MANETs have attracted a lot of attention. In these networks, each node can transmit data directly to nodes within its transmission range. To communicate with distant nodes, cooperation of intermediate nodes is essential.

Many existing protocols have taken it for granted that all nodes are cooperative. This assumption, however, is reasonable only in situations that all mobile nodes belong to the same authority and share a common goal (such as military situations). Since cooperation will incur costs to a node, in many applications nodes are not willing to cooperate unless they benefit from participation. These nodes, who seek to maximize their profit, are called \emph{selfish} nodes. In ad hoc routing, for instance, a selfish node may choose not to forward packets for other nodes so as to save limited resources, such as battery power.

To cope with the selfish behaviors, two general approaches have been proposed; namely reputation based methods and pricing based methods. In reputation based methods \cite{rep:1,rep:2,rep:3}, the behavior of nodes are monitored in order to punish non-cooperative nodes. Pricing based methods \cite{ADHOCVCG:4,CrypVCG:5,IPASS:6,Lowcost:7,SPRITE:8,OURS:9,collusion:10,Dynamic:11,Beysian:12,liu2010:22,liu2007:23,Commit:24}, however, take a proactive approach. They provide incentive for selfish nodes to act cooperatively by paying them in compensation. Although these approaches have been studied widely, there are several key issues to be considered.

One critical issue which has been neglected in existing methods, is the mobility of nodes. In MANETs, as its name suggests, nodes are mobile. Therefore, link breakage between neighbors occurs repeatedly, which in turn may lead to significant performance degradation in such protocols~\cite{khaledi:3,khaledi:2}. To cope with this, several mobility aware protocols have been proposed \cite{Gerlapredic:13,RORP:14, khaledi:1}. In these protocols, mobility prediction methods are utilized to predict durations of each route. In this way, the adverse effects of node mobility can be minimized. For example, in ad hoc routing, by predicting path durations, more stable routes can be chosen for data transmission. Consequently, number of route discoveries, which are accompanied by some delay and overhead, are reduced. However, none of these protocols consider selfish behaviors. In fact, they assume that all nodes act cooperatively.

Furthermore, most current pricing based methods pay selfish nodes considerably more than their actual costs of participation. In order to stimulate nodes to cooperate, paying premiums over their actual costs is inevitable. However, the sum of these premiums, called \emph{overpayments}, can raise uncontrollably. Therefore, it is desirable to reduce overpayments as much as possible.

Motivated by the above-mentioned issues, in this paper we propose a pricing based method for mobility aware routing in MANETs. Our method is based on the multi-dimensional optimal auctions \cite{Multidimensionalauction:16} which allows us to consider path durations in addition to route costs, when selecting the winner route. Moreover, the proposed mechanism minimizes the total expected payment over all mechanisms.

The contributions of this paper are as follows. First, we introduce a multi-dimensional auction to stimulate cooperation among the selfish nodes. The proposed auction uses \emph{Bayesian Nash equilibrium} which ensures truthful bidding. In this auction, each bid consists of two values: predicted path duration of the route and its cost of forwarding packets. Since the auctioneer (destination node) knows the path duration of the winner route, it can hold the auction again just before the current route breaks. Therefore, by using the predicted path durations in adaptive holding of auctions, the unpleasant effects of mobility can be reduced. Furthermore, holding auction between different routes helps us to pay fewer premiums. To the best of our knowledge, this is the first multi-dimensional mechanism introduced to provide incentive in MANETs. Moreover, none of the pricing based methods have considered mobility aware routing. Second, the proposed mechanism is optimal with regard to the payments \cite{Optimalmeyerson:17}. Owing to the selfish nature of nodes, providing incentives will cost more than actual costs incurred by the packet forwarding. We optimize the total expected payments in our proposed auction. In other words, the sum that auctioneer has to pay is minimized. Third, a profit sharing mechanism is proposed to ensure cooperation among the nodes of the winner route.

The rest of this paper is organized as follows. Section \ref{relwork} provides a categorization along with the review of existing methods. In addition, we present some relevant concepts in this section. Section \ref{sysmodel} is devoted to the system model used in this paper. In section \ref{propmethod}, we present the proposed method in detail. Simulation results are reported in section \ref{sim}. Finally, the concluding remarks are provided in section \ref{conclue}.



\section{Related work}\label{relwork}
Mechanism design deals with the design of systems such that the agents selfish behaviors yield the desired outcome \cite{Nisan:19}. Based on this concept, many pricing based methods have been proposed to stimulate cooperation in the MANET routing protocols. These methods can be categorized according to their solution concepts. The first category of methods uses a strong concept called the \emph{dominant equilibrium} \cite{ADHOCVCG:4,CrypVCG:5,IPASS:6,Lowcost:7,SPRITE:8,Commit:24}. In dominant equilibrium, each node selects its best strategy regardless of other nodes strategies. The second category implements a weak solution concept called the \emph{Nash equilibrium} \cite{OURS:9,collusion:10,liu2010:22,liu2007:23}. In Nash equilibrium, given the other nodes strategies, each node chooses its best action. The third category uses the Bayesian Nash equilibrium \cite{Dynamic:11,Beysian:12} which is less stringent than the dominant equilibrium but stronger than the Nash equilibrium. Each node in the Bayesian Nash equilibrium chooses a strategy with the maximum expected utility, given the distribution of nodes.

In general, there are some desirable properties for pricing based mechanisms which can be summarized in the followings:
\begin{itemize}
\item \emph{Individual Rationality (IR)}: a mechanism is called individually rational if utility of each node is always nonnegative. Otherwise, nodes may choose not to participate.

\item \emph{Incentive Compatibility (IC)}: In an incentive compatible mechanism, no selfish node has incentive to lie.

\item \emph{Budget Balance (BB)}: a budget balanced mechanism pays nodes exactly their actual costs and nothing more.

\item \emph{Collusion Resistance (CR)}: a mechanism is called collusion resistant if no colluding subset of nodes, can increase the sum of their profit by misreporting.

\item \emph{Mobility Awareness (MA)}: due to mobile nature of MANETs, a mechanism should consider changes made by mobility of nodes.
\end{itemize}

\begin{table}[!t]
\renewcommand{\arraystretch}{1.4}
\caption{The categories of pricing based methods.'C' means Conditional.} \label{tblrelwork} \centering
\begin{tabular}{l l c c c c c}
\hline\hline
Category& Approach& IR & IC & BB & CR & MA \\
\hline
\multirow{3}{*}{Dominant} & Ad Hoc-VCG\cite{ADHOCVCG:4} & \checkmark& \checkmark& $\times$& $\times$& $\times$\\
 & Low cost\cite{Lowcost:7} & \checkmark& \checkmark& $\times$& $\times$& $\times$\\
 & Sprite\cite{SPRITE:8} &\checkmark &C& $\times$&C &$\times$\\\hline

 \multirow{2}{*}{Nash} & Collusion Resistant\cite{collusion:10}& \checkmark&\checkmark &$\times$&\checkmark &$\times$\\
 & OURS\cite{OURS:9} &\checkmark &\checkmark &$\times$& $\times$&$\times$\\\hline

 \multirow{3}{*}{Bayesian Nash} & Dynamic Pricing\cite{Dynamic:11}&\checkmark &\checkmark & $\times$& $\times$&$\times$\\
 & BIC-B\cite{Beysian:12} & C& \checkmark& \checkmark& $\times$&$\times$ \\
 & The Proposed Method &\checkmark &\checkmark& $\times$&$\times$ & \checkmark\\\hline\hline
\end{tabular}
\end{table}

Table \ref{tblrelwork} shows three categories of existing pricing based methods and the properties of notable papers in each category. This categorization which is presented for the first time in literature, demonstrates the current state of research in this field.

One of the most prominent papers in dominant category is Ad Hoc-VCG \cite{ADHOCVCG:4} which makes use of the celebrated VCG mechanism. Although Ad Hoc-VCG provides incentive compatibility in dominant equilibrium, it requires enormous payments to selfish nodes. Furthermore, collusion of nodes and their mobility is not considered in Ad Hoc-VCG. Also, some of its underlying assumptions are problematic. For example, the mechanism designer must be aware of the network topology and the network graph must be two-connected. In \cite{Lowcost:7}, authors propose an algorithm to implement VCG in a distributed manner. However, they have ignored the imposed network complexity. Sprite \cite{SPRITE:8} is another pricing based system which utilizes cryptographic techniques to ensure cooperation among selfish nodes. It provides collusion resistance and incentive compatibility in a conditional manner. In essence, Sprite is a receipt-submission game which is best suited for unicast routing.

By using the Nash equilibrium solution concept, there have been some efforts to overcome shortcomings of VCG. In \cite{OURS:9}, for instance, authors propose a method to reduce large overpayments of VCG. Also a collusion resistant mechanism is proposed in \cite{collusion:10}. Authors apply cryptographic techniques to prevent profit transfer among colluding nodes. An underlying assumption in Nash equilibrium is that each selfish node selects its strategy with knowing others' strategies. However, due to mobile and multihop nature of MANETs, this assumption seems unrealistic.

Because of drawbacks of the Nash equilibrium, some methods have recently been proposed based on the Bayesian Nash equilibrium as the solution concept. For instance, Suri and Narahari \cite{Beysian:12} proposed a budget balanced mechanism, called BIC-B, for broadcasting. BIC-B avoids any overpayments. However, this may violate individual rationality. Without being individually rational, nodes may choose not to participate in protocol. Authors determined a condition under which individual rationality holds. In \cite{Dynamic:11}, authors model the MANET routing problem as a multi-stage pricing game. Each stage corresponds to one route discovery phase in ad hoc on-demand routing. Their proposed mechanism seeks to maximize the sender-receiver's payoff by varying number of transmitting packets with respect to the cost of chosen route. That is, more packets will be transmitted at stages having lower cost compared to the previous stages. Although their proposed approach maximizes the sender-receiver payoff by adaptively adjusting the transmission rate at different stages, it is prone to frequent route breakages caused by mobility of nodes. This happens because only costs of routes are considered in choosing the winner route. As a result of ignoring stability of routes, more packets will get lost and route discovery will be repeated numerously. The situation gets even worse if transmission rate is increased just because of a lower cost.

As mentioned above, mobility of nodes has not been considered in the previous works. In other words, they have only considered cost of forwarding packets. In this paper, we propose a multi-dimensional mechanism which uses both the cost and the predicted path duration of routes. This mechanism ensures cooperation among the selfish nodes in mobility aware routing protocols and has the ability to cope with frequent topology changes caused by mobility. Also, the proposed mechanism provides Individual Rationality and avoids large overpayments. Moreover, our proposed mechanism is not required to impose restrictive assumptions like those of the Ad Hoc-VCG.

\section{System Model}\label{sysmodel}

We consider routing problem for MANETs in presence of selfish nodes. All nodes are rational and selfish; each may belong to a different authority. Their main goal is to maximize their profit, not to harm others. Since forwarding packets for others will incur some cost, a selfish node may not be willing to participate in routing. Therefore, sufficient incentives should be provided for nodes.

In this paper, we consider reactive routing protocols that handle the routing in two phases; route discovery and data transmission. The source node establishes a route to the destination in route discovery phase, when it has packets to send. Then, data is transmitted in the data transmission phase. Due to mobile nature of MANETs, a source node needs to perform route discovery frequently. To deal with this issue, several mobility aware routing protocols have been proposed \cite{Gerlapredic:13,RORP:14}. Using mobility prediction methods, these protocols have the ability to cope with frequent topology changes. Therefore, with less route breakage, number of route discovery phases will be reduced. As a result, higher throughput and less overhead can be achieved in comparison with non-mobility aware protocols.

We use multi-dimensional mechanism design to model the routing procedure. In a mechanism, there are $n$ agents; each has some private value $t_i$, called its \emph{type}. According to the revelation principle \cite{Nisan:19}, we can restrict our attention to direct revelation mechanisms in which the reported value is the agent's type. Each agent's type consists of cost of forwarding, $c$, and predicted path duration of the route, $d$. A type vector $t=(t_1,t_2,\ldots,t_n)$ comprising all agents' types is called \emph{profile}. Given a reported profile, the mechanism determines the output as well as the payments to each agent. Each agent's utility is defined as its total received payment minus its cost of participation.

In the routing game, Bayesian Nash equilibrium is used as a solution concept. In this setting, it is assumed that every player has common knowledge of the distribution of other players' type. Each route forms a single bid, (c, d), and announces it's cost and path duration. Then, the winner route and the total payment are determined by the destination node. In addition, the obtained profit is shared among the winner route nodes. For profit sharing, we used a simple game which is based on nash equlibrium. In this game, it is assumed that nodes of the winner route are aware of costs of forwarding and the total payment.

\section{The proposed method}\label{propmethod}
In this paper, we propose a multi-dimensional optimal auction to provide a game theoretical framework in mobility aware routings. The proposed auction implements Bayesian Nash equilibrium which makes more reasonable assumptions about node information than the Nash equilibrium. The destination node holds the auction and determines the winner route between a source and a destination node. Like Ji \emph{et. al} \cite{Dynamic:11}, our auction is held between different routes in the route discovery phase. However, our proposed auction is multi-dimensional. That is, type of each bidder consists of two values, namely the cost of forwarding $c_i$ and the predicted path duration of the route, $d_i$. Cost of each route is the sum of its belonging nodes' costs and Path duration of a route is the minimum of its link durations. In route discovery phase, each node appends its own cost and predicted link duration to the route request packet. Thus, the destination node receives a list of costs and link durations for each path (that form a single bid). Then, the destination node holds the auction based on received bids. It is important to note that by considering paths as bidding entities fewer premiums are paid. For example, consider a network with two node-disjoint paths, $P_1$ and $P_2$, between a source-destination pair. Suppose that each node on $P_1$ has cost zero and each node on $P_2$ has cost one. Thus, total cost of $P_1$ is zero and total cost of $P2$ is $N$ (assuming $N$ nodes on each path). Using VCG mechanism, the auctioneer has to pay an overpayment equal to $N^2$ which is more expensive than the second cheapest path ($P_2$). While, in case that each path is a bidder, overpayment of VCG mechanism equals $N$ \cite{Nisan:19}.

In order to predict link durations, we utilize a widely used prediction method described in \cite{Gerlapredic:13}. According to \cite{Gerlapredic:13}, each node transmits its mobility information (its coordinates, moving speed and direction) to its neighbor via route request packet. Then, the neighbor predicts the amount of time that these two nodes will stay connected (the link duration) using a simple formula.

The auction is described by a pair of functions $(A,P)$ , namely the allocation function and the payment function. $A_{c,d}$ shows the probability that $(c,d)$ wins and $p_{c,d}$ specifies the expected payment to $(c,d)$. Since the auctioneer knows path duration of the winner route, it can hold the auction again just before the current route breaks.

Let $f_c$, $f_d$ represent probability of having cost of $c$ and path duration of $d$ respectively. Since forwarding cost of a route is independent from its path duration, $f_{c,d}=f_c\times f_d$ will be the probability that a bidder has type $(c,d)$. Note that, we assume independent and discrete types.

Let $t^{n-1}$ denote the profile involving only $n-1$ agents, which is realized with the probability of $\Pi(t^{n-1})$. Also, let $w(A_{c,d}[t^{n-1}]|c,d)=c A_{c,d}$ be the $(c,d)$'s cost regarding the allocation probability $A_{c,d}$. Then, the expected cost of $(c,d)$ with respect to $A_{c,d}$ and its expected utility, $u_{c,d}$, will be:

\begin{equation}\label{equexpcost}
E_{t^{n-1}}[w(A_{c,d}[t^{n-1}]|c,d)]=\sum_{t^{n-1}}\Pi(t^{n-1})w(A_{c,d}[t^{n-1}]|c,d)
\end{equation}
\begin{displaymath}\label{equexputility}
u_{c,d}=p_{c,d}-E_{t^{n-1}}[w(A_{c,d}[t^{n-1}]|c,d)]
\end{displaymath}

Define $a_{c,d}$ as the following:
\begin{displaymath}\label{equa}
a_{c,d}=\sum_{t^{n-1}}\Pi(t^{n-1})A_{c,d}[t^{n-1}]
\end{displaymath}
Then, (\ref{equexpcost}) can be rewritten as:
\begin{equation}\label{equsexpcost}
w(a_{c,d}|c,d)=ca_{c,d}.
\end{equation}
In this context, we need some definitions. \emph{Bayesian incentive compatibility} which is required to ensure that bidders report truthfully, means:
\begin{equation}\label{equic}
\forall\:(c,d);\forall\:(c',d'):\:p_{c,d}-w(a_{c,d}|c,d)\geq p_{c',d'}-w(a_{c',d'}|c,d)
\end{equation}
Where $(c',d')$ represent the misreported type of the agent with type $(c,d)$. A mechanism is called \emph{individually rational} if:
\begin{displaymath}\label{equir}
\forall\:(c,d):\:p_{c,d}-w(a_{c,d}|c,d)\geq 0
\end{displaymath}
An allocation rule is \emph{monotonic} if:
\begin{equation}\label{equmonotone}
\forall\:(c,d);\forall\:(c',d'):\:(c,d)\succeq (c',d')\Rightarrow a_{c,d}\geq a_{c',d'}
\end{equation}
where $(c,d)\succeq(c',d')$ means that $(c,d)$ is partially ordered above $(c',d')$. A type is partially ordered above the other, if it has a lower cost and a greater path duration. We need to define another condition called \emph{Monotone Hazard rate} which states:
\begin{equation}\label{equmonotonehazardrate}
\forall\:(c,d);\forall\:(c',d'):\:(c,d)\succeq (c',d')\Rightarrow \frac{f_{c,d}(c|d)}{1-F_{c,d}}\geq \frac{f_{c',d'}(c'|d')}{1-F_{c',d'}}
\end{equation}
In the above condition, $f_{c,d}(c|d)$ denotes the probability density function of $(c,d)$ conditional on $d$ and $F_{c,d}=\sum_{k=c+1}^{c_{max}}f_{k,d}(k|d)$. The monotone hazard rate condition holds for many distributions like uniform, exponential, normal, pareto, etc.

\emph{Bayesian Nash Revelation Principle} states that if there is a mechanism (direct or otherwise) that implements the social-choice function $f$ in Bayesian Nash equilibrium, then $f$ is truthfully implementable in a Bayesian incentive compatible direct-revelation mechanism \cite{Nisan:19}. In other words, in order to implement a particular social choice function in Bayesian Nash equilibrium, it is sufficient to consider only incentive compatible direct-revelation mechanisms. Therefore, we only prove Bayesian incentive compatibility of our mechanism.

Owing to numerous cases formed by combinations of cheating in each dimension, with multi-dimensional types, proving incentive compatibility is not a trivial task. In subsection A, we mention and simplify incentive compatibility constraints. In subsection B and subsection C we introduce the optimal payment function and the optimal allocation rule that satisfy incentive compatibility, respectively. Finally, in subsection D, a mechanism is proposed to share obtained profit among nodes of the route.

\subsection{Incentive Compatibility Constraints}\label{subsecic_reduc}
Bidders may cheat in reporting their values in order to gain some extra profit. Incentive Compatibility (IC) ensures that agents have no incentive to misreport their values. In our setting, a bidder can report its cost and path duration more or less than their actual values.

As we have considered Individual Rationality (IR) constraint, IC constraint for over-reporting of path duration is redundant. Suppose a bidder with type $(c,d)$ misreports its type as $(c,d')$, where $d' > d$. IC constraint implies:
\begin{displaymath}
p_{c,d}-w(a_{c,d}|c,d)\geq p_{c,d'}- w(a_{c,d'}|c,d)
\end{displaymath}
Regarding the point that payments are made at the end of data transmission phase, if a bidder over-reports its path duration, it will get its expected cost that means $P_{c,d'}=c a_{c,d}$ because it is not present. Therefore:

\begin{displaymath}
p_{c,d}-w(a_{c,d}|c,d)\geq c(a_{c,d}-a_{c,d'})
\end{displaymath}

According to the Individual Rationality and the monotonicity of the allocation rule, the above inequality holds. As a result, IC constraint for over-reporting the path duration can be excluded, and the remaining IC constraints are as follows:

\begin{enumerate}
\item $(c,d) \rightarrow (c',d)$ where $c>c'$
\item $(c,d) \rightarrow (c',d)$ where $c<c'$
\item $(c,d) \rightarrow (c,d')$ where $d>d'$
\item $(c,d) \rightarrow (c',d')$ where $c>c'$ and $d>d'$
\item $(c,d) \rightarrow (c',d')$ where $c<c'$ and $d>d'$
\end{enumerate}
Where $(c,d) \rightarrow (c',d)$ states that agent with type $(c,d)$ must not benefit from untruthfully reporting $(c',d)$. In order to simplify the above constraints, we first introduce some Lemmas.

\newtheorem{lemma}{Lemma}
\begin{lemma}\label{lem1}
IC constraints 4 and 5 are redundant if constraints 1, 2 and 3 hold.
\end{lemma}
\begin{IEEEproof}[Proof]
We show that constraint 4 can be implied by applying constraint 3 followed by constraint 1.
According to (\ref{equic}), adding IC constraints 1 and 3 yields:
\begin{displaymath}\label{lem1:1}
\begin{split}
p_{c,d}-w(a_{c,d}|c,d)-w(a_{c,d'}|c,d')\geq &\:p_{c',d'}-w(a_{c,d'}|c,d)-w(a_{c',d'}|c,d')
\end{split}
\end{displaymath}
By adding $w(a_{c',d'}|c,d)$ to both sides of the above inequality, we have:
\begin{displaymath}\label{lem1:2}
\begin{split}
[p_{c,d}-p_{c',d'}]&-[w(a_{c,d}|c,d)-w(a_{c',d'}|c,d)]\geq w(a_{c,d'}|c,d')\\
&-w(a_{c,d'}|c,d)-w(a_{c',d'}|c,d')+w(a_{c',d'}|c,d)
\end{split}
\end{displaymath}
By Lemma 5 proposed in Appendix B, the allocation rule is monotonic. According to definition of $w$, (\ref{equsexpcost}), and monotonicity of allocation rule, (\ref{equmonotone}), the right hand side of the above inequality is greater than or equal to zero. Therefore:
\begin{displaymath}
\label{lem1:3}
[p_{c,d}-p_{c',d'}]\geq [w(a_{c,d}|c,d)-w(a_{c',d'}|c,d)]
\end{displaymath}
Which is the IC constraint 4. By a similar approach, constraint 5 can be obtained by constraints 2 and 3.
\end{IEEEproof}

In the next Lemma, we show that only the adjacent IC constraints matter. In other words, cases in which an agent misreports its value two or more units above (or below) its real value are implied by cases of incrementing (or decrementing) its actual value by one unit.

\begin{lemma}\label{lem2}
Only adjacent IC constraints must be considered.
\end{lemma}
\begin{IEEEproof}[Proof]
We prove this Lemma for the case that an agent over-reports its cost. The case of underreporting cost or path duration can be proved similarly.
It suffices to show that the following pair of constraints:
\begin{flalign}
&p_{c,d}-w(a_{c,d}|c,d)\geq p_{c+1,d}-w(a_{c+1,d}|c,d)\label{equlem21}\\
&p_{c+1,d}-w(a_{c+1,d}|c+1,d)\geq p_{c+2,d}-w(a_{c+2,d}|c+1,d)\label{equlem22}
\end{flalign}
Imply:
\begin{displaymath}
p_{c,d}-w(a_{c,d}|c,d)\geq p_{c+2,d}-w(a_{c+2,d}|c,d)
\end{displaymath}

Then the rest will follow by induction. By adding (\ref{equlem21}) and (\ref{equlem22}), we have:
\begin{displaymath}\label{lem2:2}
\begin{split}
p_{c,d}-w(a_{c,d}|&c,d)-w(a_{c+1,d}|c+1,d)\geq p_{c+2,d}
-w(a_{c+1,d}|c,d)-w(a_{c+2,d}|c+1,d)
\end{split}
\end{displaymath}
Rearranging and adding $w(a_{c+2,d}|c,d)$ to both sides of the inequality, yields:
\begin{displaymath}\label{lem2:3}
\begin{split}
[p_{c,d}-p_{c+2,d}]&-[w(a_{c,d}|c,d)-w(a_{c+2,d}|c,d)]\geq \\
& w(a_{c+1,d}|c+1,d)-w(a_{c+1,d}|c,d)
-w(a_{c+2,d}|c+1,d)+w(a_{c+2,d}|c,d)
\end{split}
\end{displaymath}
According to definition of $w$, (\ref{equsexpcost}), and monotonicity of allocation rule, (\ref{equmonotone}), the right hand side of the above inequality is greater than or equal to zero. Therefore:
\begin{displaymath}\label{lem2:4}
[p_{c,d}-p_{c+2,d}]\geq [w(a_{c,d}|c,d)-w(a_{c+2,d}|c,d)]
\end{displaymath}
\end{IEEEproof}
With the aid of Lemma \ref{lem1} and Lemma \ref{lem2}, IC constraints are reduced in the following Theorem.
\newtheorem{theorem}{Theorem}
\begin{theorem}\label{The1}
All IC constraints are implied by the followings:
\begin{equation}\label{equthe1}
\begin{split}
(c, d)\rightarrow(c+1, d)\\
(c, d)\rightarrow(c-1, d)\\
(c, d)\rightarrow(c, d-1)
\end{split}
\end{equation}
\end{theorem}
\begin{IEEEproof}[Proof]
Using Lemma \ref{lem1}, only IC constraints 1, 2, 3 are needed to be considered. Also, Lemma \ref{lem2} shows that only adjacent IC constraints matter. Therefore, all IC constraints are implied by (\ref{equthe1}).
\end{IEEEproof}

\subsection{The Optimal Payment Function}\label{sbsecoptimalprice}
Here we suppose a monotonic allocation rule is given, and then we characterize the optimal payment function. In the next subsection, we characterize the allocation rule with respect to the specified payment rule. The goal is to find a payment rule which minimizes the sum that auctioneer has to pay. The problem can be stated as follows:
\begin{figure}[!t]
\centering
\includegraphics[width=8.5cm,height=8cm]{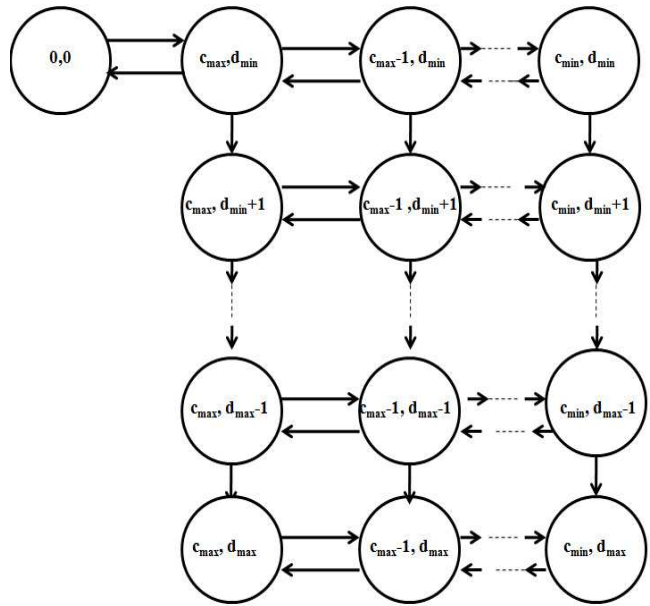}
\caption{The network graph.} \label{netgraph}
\end{figure}
\begin{gather}
\min_p{\sum_d\sum_cf_{c,d}\:p_{c,d}}\label{equoptimizeprice}\\
s.t.\nonumber\\
p_{c,d}-w(a_{c,d}|c,d)\geq p_{c+1,d}-w(a_{c+1,d}|c,d)\label{optpricecons1}\\
p_{c,d}-w(a_{c,d}|c,d)\geq p_{c-1,d}-w(a_{c-1,d}|c,d)\label{optpricecons2}\\
p_{c,d}-w(a_{c,d}|c,d)\geq p_{c,d-1}-w(a_{c,d-1}|c,d)\label{optpricecons3}\\
p_{c,d}-w(a_{c,d}|c,d)\geq 0\label{optpricecons4}
\end{gather}

Constraint (\ref{optpricecons4}) is imposed by individual rationality (IR) and (\ref{optpricecons1}), (\ref{optpricecons2}) and (\ref{optpricecons3}) correspond to IC constraints determined by Theorem \ref{The1}. Based on \cite{Multidimensionalauction:16}, this problem is the dual of finding the longest path in a graph. Consider a graph in which each vertex of the graph represents a type. For each IC constraint we define an edge with length specified as follows:

\begin{itemize}
\item $(c, d)\rightarrow(c+1, d)$: an edge from $(c+1, d)$ to $(c, d)$ with length $w(a_{c,d}|c,d)-w(a_{c+1,d}|c,d)=c(a_{c,d}-a_{c+1,d})$
\item $(c, d)\rightarrow(c-1, d)$: an edge from $(c-1, d)$ to $(c, d)$ with length $w(a_{c,d}|c,d)-w(a_{c-1,d}|c,d)=c(a_{c,d}-a_{c-1,d})$
\item $(c, d)\rightarrow(c, d-1)$: an edge from $(c, d-1)$ to $(c, d)$ with length $w(a_{c,d}|c,d)-w(a_{c,d-1}|c,d)=c(a_{c,d}-a_{c,d-1})$
\end{itemize}

We also introduce a dummy vertex corresponding to a dummy type $(0,0)$. Payment to the dummy type, as well as its allocation probability, is zero. This enables us to consider IR constraint as an extra IC constraint. The length of edge from the dummy node to $(c_{max},d_{min})$ equals to $c_{max}a_{c_{max},d_{min}}$. Fig. \ref{netgraph} depicts the resulting graph.

The optimal payment to $(c,d)$, $p_{c,d}$, corresponds to the length of the longest path from the dummy node $(0,0)$ to vertex $(c,d)$. For the sake of solvability, there must be no positive cycles in the graph. As Fig. \ref{netgraph} illustrates, it suffices to show that there are no positive cycles between neighboring nodes. For instance, consider $(c,d)$ and $(c+1,d)$, the following condition must hold:
\begin{gather}
[w(a_{c,d}|c,d)-w(a_{c+1,d}|c,d)]+[w(a_{c+1,d}|c+1,d)-w(a_{c,d}|c+1,d)]\leq 0\nonumber\\
c(a_{c,d}-a_{c+1,d})+(c+1)(a_{c+1,d}-a_{c,d})\leq 0\nonumber\\
(-a_{c,d}+a_{c+1,d})\nonumber\leq 0
\end{gather}

According to the monotonicity: $a_{c+1,d}\leq a_{c,d}$. Therefore, there is no positive cycle between $(c,d)$ and $(c+1,d)$.

\begin{theorem}\label{The2}
Given a monotonic allocation rule, the optimal payment rule is:

\begin{equation}\label{optprice}
p_{c,d}=ca_{c,d}+\sum_{k=c+1}^{c_{max}} a_{k,d}
\end{equation}

\end{theorem}
\begin{IEEEproof}[Proof]
Let $IC_d$ and $IC_u$ be the IC constraints when a node underreports and overreports its value respectively. We remove edges corresponding to $IC_d$ constraints from the graph. In Fig. \ref{netgraph}, these constraints correspond to the links connecting vertices on the right to the left-hand vertices. In the Appendix A section, we prove the redundancy of $IC_d$ constraints.
As stated earlier, $p_{c,d}$, equals to the length of the longest path from $(0,0)$ to $(c,d)$. In order to prove (\ref{optprice}), we show that the longest path is the one that first goes down the bottom, then moves to the side. In other words, the longest path to $(c,d)$ is of the following form:
\begin{flalign*}
&(0,0)\longrightarrow(c_{max},d_{min})\longrightarrow(c_{max},d_{min}+1)\longrightarrow(c_{max},d_{min}+2)\ldots\longrightarrow(c_{max},d)\longrightarrow\\
&(c_{max}-1,d)\longrightarrow(c_{max}-2,d)\ldots\longrightarrow(c,d)
\end{flalign*}
For $(c_{max}-1,d_{min})$ and $(c_{max},d_{min}+1)$, only one path exists. Therefore:
\begin{flalign*}
&p_{c_{max}-1,d_{min}}=a_{c_{max},d_{min}}+(c_{max}-1)a_{c_{max}-1,d_{min}}\\
&p_{c_{max},d_{min}+1}=c_{max}a_{c_{max},d_{min}+1}
\end{flalign*}
Now, suppose that we wish to determine  $p_{c_{max}-1,d_{min}+1}$ . There are two possible paths:
\begin{flalign*}
&P_1:(0,0)\longrightarrow(c_{max},d_{min})\longrightarrow(c_{max}-1,d_{min})\longrightarrow(c_{max}-1,d_{min}+1)\\
&P_2:(0,0)\longrightarrow(c_{max},d_{min})\longrightarrow(c_{max},d_{min}+1)\longrightarrow(c_{max}-1,d_{min}+1)
\end{flalign*}
With lengths:
\begin{flalign*}
P_1:& c_{max}a_{c_{max},d_{min}}+(c_{max}-1)(a_{c_{max}-1,d_{min}}-a_{c_{max},d_{min})}+(c_{max}-1)(a_{c_{max}-1,d_{min}+1}-\\
&a_{c_{max}-1,d_{min}})=a_{c_{max},d_{min}}+(c_{max}-1)a_{c_{max}-1,d_{min}+1}
\end{flalign*}
\begin{flalign*}
P_2:& c_{max}a_{c_{max},d_{min}}+c_{max}(a_{c_{max},d_{min}+1}-a_{c_{max},d_{min}})+(c_{max}-1)(a_{c_{max}-1,d_{min}+1}-\\
&a_{c_{max},d_{min}+1})=a_{c_{max},d_{min}+1}+(c_{max}-1)a_{c_{max}-1,d_{min}+1}
\end{flalign*}

Using monotonicity of the allocation rule, (\ref{equmonotone}), $a_{c_{max},d_{min}+1} \geq a_{c_{max},d_{min}}$; thus $P_2$ is longer than $P_1$. Assume that for all types $(a,b)$ where $a\geq c_{min}+1$ and $b\leq d_{max}-1$, the longest path is of the claimed form. We will show that the longest paths for all $(a,d_{max}),\: a\geq c_{min}+1$ and for all $(c_{min},b),\: b\leq d_{max}-1$, are obtained by first going down the bottom, then moving to the side.

Clearly, our claim is true for $(c_{max},d_{max})$. Also, for $(c_{max}-1,d_{max})$ an almost identical argument to $(c_{max}-1,d_{min}+1)$ can be applied. Now consider $(c_{max}-2,d_{max})$, there are two possible paths:
\begin{flalign*}
&P_1:(0,0)\longrightarrow(c_{max},d_{max}-1)\longrightarrow(c_{max}-2,d_{max}-1)\longrightarrow(c_{max}-2,d_{max})\\
&P_2:(0,0)\longrightarrow(c_{max},d_{max}-1)\longrightarrow(c_{max},d_{max})\longrightarrow(c_{max}-2,d_{max})
\end{flalign*}
With lengths:
\begin{flalign*}
P_1:& p_{c_{max},d_{max}-1}+(c_{max}-2)(a_{c_{max}-2,d_{max}-1}-a_{c_{max},d_{max}-1})+(c_{max}-2)(a_{c_{max}-2,d_{max}}-\\
&a_{c_{max}-2,d_{max}-1})=p_{c_{max},d_{max}-1}+(c_{max}-2)(a_{c_{max}-2,d_{max}}-a_{c_{max},d_{max}-1})
\end{flalign*}
\begin{flalign*}
P_2:& p_{c_{max},d_{max}-1}+c_{max}(a_{c_{max},d_{max}}-a_{c_{max},d_{max}-1})+(c_{max}-2)(a_{c_{max}-2,d_{max}}-a_{c_{max},d_{max}})\\
&=p_{c_{max},d_{max}-1}+2a_{c_{max},d_{max}}+(c_{max}-2)a_{c_{max}-2,d_{max}}-c_{max}a_{c_{max},d_{max}-1}
\end{flalign*}
Once again, using monotonicity of allocation rule, $ P_2$ is longer than $P_1$. Similarly, the longest paths to $(a,d_{max})$ and $(c_{min},b)$ where $a\geq c_{min}+1$ and $b\leq d_{max}-1$, follow the claimed form. The only remaining case is for $(c_{min},d_{max})$:
\begin{flalign*}
&P_1:(0,0)\longrightarrow(c_{max},d_{max}-1)\longrightarrow(c_{min},d_{max}-1)\longrightarrow(c_{min},d_{max})\\
&P_2:(0,0)\longrightarrow(c_{max},d_{max}-1)\longrightarrow(c_{max},d_{max})\longrightarrow(c_{min},d_{max})
\end{flalign*}
Using monotonicity for comparing lengths of the above paths, reveals that the longest path to $(c_{min},d_{max})$ is also of the claimed form. Therefore, $p_{c,d}$ will be:
\begin{gather*}
p_{c_{max},d+1}=c_{max}a_{c_{max},d+1},\\
p_{c_{max}-1,d+1}=(c_{max}-1)a_{c_{max}-1,d+1}+a_{c_{max},d+1},\\
p_{c_{max}-2,d+1}=(c_{max}-2)a_{c_{max}-2,d+1}+a_{c_{max}-1,d+1}+a_{c_{max},d+1}, \ldots\\
p_{c,d}=ca_{c,d}+\sum_{k=c+1}^{c_{max}} a_{k,d}
\end{gather*}
\end{IEEEproof}
Note that by Theorem 4 proposed in Appendix B, the optimal payment function meets monotonicity and IC.

\subsection{The Optimal Allocation Rule}\label{subsecoptimalalloc}
In the previous section, we characterized the payment rule assuming that a monotonic allocation rule is given. In this section, we specify the allocation rule and show that it is monotonic. Recall that we stated the problem in (\ref{equoptimizeprice}). Having the optimal price, characterized in Theorem \ref{The2}, the problem can be rewritten as:

\begin{gather}
\min_a{\sum_d\sum_cf_{c,d}[ca_{c,d}+\sum_{c+1}^{c_{max}}a_{k,d}]}\nonumber\\
s.t.\nonumber\\
\forall\:(c,d);\forall \:(c',d'):\:(c,d)\succeq (c',d')\Rightarrow a_{c,d}\geq a_{c',d'}\label{equoptall1}\\
\forall \:(c,d):\:a_{c,d}=\sum_{t^{n-1}}\Pi(t^{n-1})A_{c,d}[t^{n-1}]\nonumber\\
\forall \:t:\:\sum_d\sum_c A_{c,d}(t)\leq1\label{equoptall2}
\end{gather}

Constraint (\ref{equoptall1}) corresponds to the monotonicity of allocation rule. Also, constraint (\ref{equoptall2}) states that sum of allocation probabilities of all types must equal 1. Let  $F_{c,d}=\sum_{k=c+1}^{c_{max}}f_{k,d}(k|d)$, the above problem can be rewritten as:

\begin{gather}
\min_a{\sum_d\sum_cf_{c,d}a_{c,d}vv_{c,d}}\nonumber\\
vv_{c,d}=c+\frac{1-F_{c,d}}{f_{c,d}(c|d)}\label{equvv}
\end{gather}

Where $vv_{c,d}$ is the $(c,d)$'s virtual valuation \cite{Optimalmeyerson:17}. This is an instance of knapsack problem which can be solved in the following way. Given the distribution of types, the auctioneer can compute virtual valuation of each type. Within each profile, the type with minimum virtual valuation gets the allocation probability of $1$, and the winner will be the type with the highest expected allocation probability, $a_{c,d}$. In the next Theorem, we show that this allocation rule is monotonic.

\begin{theorem}\label{The3}

Assuming a monotone hazard rate, the proposed allocation rule is monotonic.

\end{theorem}
\begin{IEEEproof}[Proof]
According to (\ref{equvv}) and the monotone hazard rate condition, if type $(c,d)$ is partially ordered above type $(c',d')$, its virtual valuation must be less than $(c',d')$'s  virtual valuation. Therefore, with the proposed allocation rule, $(c,d)$ becomes the winner with a higher probability compared to $(c',d')$. Thus:
\begin{displaymath}
(c,d)\succeq(c',d')\Rightarrow a_{c,d}\geq a_{c',d'}
\end{displaymath}


\end{IEEEproof}

\subsection{The Profit Sharing Mechanism}\label{subsecsharprice}

In the previous subsections, we showed that bidders have no incentive to misreport their types, $(c, d)$. Also, we specified the payment to the winner in Theorem \ref{The2}. It should be noted that, in this paper, we consider path auctions in which bidders are routes. Therefore, the remaining problem is designing a mechanism to share the obtained profit among nodes of the winner route. Nodes in the winner route may make exaggerated claims to get more than their fair shares. Thus, we need a mechanism which stimulates cooperation among the nodes of the winner route by sharing the obtained profit,Theorem \ref{The2}.

We model this situation as a game which is inspired by the scheme presented in \cite{collusion:10}. In this game, players are the nodes of the winner route. Each player is required to report its cost of participation, $c_i$. Let $P$ denote the payment that winner route receives and $c=\{c_1, c_2, \cdots, c_h\}$ be the declared profile by players. Thus, the total cost of the declared profile equals to $C=\sum_{k=1}^h c_k$. Assume that players are aware of the profile and also the $P$. Let $(c, d)$ be the winner route cost and path duration respectively. Then, payment to player $i$ is defined as:

\begin{itemize}
\item if individual rationality, $P \geq a_{c,d}C$, holds for the stated profile, pay player $i$ its claimed cost, $c_i$
\item else, pay player $i$ nothing
\end{itemize}
\begin{theorem}\label{The5}
The above scheme results in a Nash equilibrium.
\end{theorem}
\begin{IEEEproof}[Proof]
To prove this, we construct a profile $c^*$ and show that this profile forms a Nash equilibrium. At first, we set up an initial profile with real costs of players. Then, each player increases its cost by a small fixed value, $\epsilon$, only if sum of the declared costs does not exceed $P$. Otherwise, the player keeps its current value. This process repeats until no changes in the profile can be made. The resulting profile, $c^*$, forms a Nash equilibrium in the sense that no player has incentive to change its cost, given the profile. In other words, the strategy profile $c^*=\{c^*_1, c^*_2, \cdots, c^*_h\}$ is in the Nash equilibrium if for all nodes,

\begin{displaymath}\label{equenash}
u_i(c^*_i, c^*_{-i}) \geq u_i(c_i, c^*_{-i}),     \forall c_i \neq c^*_i
\end{displaymath}

Since payment to each player is equal to the claimed cost, there will be no gain in underreporting the costs, $(c_i < c^*_i)$.
Also, if the player is tempted to increase its cost and go beyond the final profile, $(c_i > c^*_i)$, it gets nothing. Thus, this scheme results in an equilibrium in which each node gets at least an amount equal to its real cost, or slightly higher than its cost.
\end{IEEEproof}

Please note that in the proposed mechanism, the auctioneer selects the winner route and computes the total payment based on route costs, the sum of nodes' costs, and path durations, minimum of link durations. In the previous subsections, we proved that our mechanism is IC with regards to route costs and path durations. The profit sharing mechanism only provides incentive for the winner route nodes to cooperate. Also, the value of link duration only has impact on the amount of expected allocation, $a_{c, d}$. Therefore, nodes of the winner route have no incentive to misreport their link durations.

\section{Simulation Results}\label{sim}

In this paper, we propose a pricing based method for mobility aware routing in MANETs that possesses some desirable properties (note that we did not aim to propose a new mobility aware routing). Performance gains of mobility aware routing have already been presented in literature (see \cite{Gerlapredic:13} for example). Mobility aware routing protocols have the ability to cope with frequent topology changes. Therefore, with less route breakage, number of route discovery phases will be reduced. As a result, higher throughput and less overhead can be achieved in comparison with non-mobility aware protocols. These performance gains lie in the area of classical routing which is not the in the scope of this paper. Thus, in this section, we only evaluate the proposed mechanism with respect to the payments.

As mentioned in the previous sections, the proposed method results in Bayesian Nash equilibrium which is weaker than the dominant equilibrium but stronger than the Nash equilibrium. In order to show the performance gains achieved by changing the solution concept from dominant equilibrium to Bayesian Nash equilibrium, we compare the proposed method with the Ad Hoc-VCG, a widely used mechanism which uses the celebrated VCG method. Ad Hoc-VCG implements incentive compatibility in dominant equilibrium with only considering the route costs and disregarding mobility of nodes. We show that our new mechanism avoids large overpayments which are common in Ad Hoc-VCG.

We consider a network in which nodes are placed in a 1000$m\times$1000$m$ area with 150$m$ transmission range for each node. Mobile nodes move according to the \emph{random walk} mobility model \cite{MobilitySurvy:21}. Since forwarding cost of a route is independent from its path duration, the probability that a bidder has type $(c,d)$ is obtained by multiplying probabilities of $c$ and $d$. Each node chooses its forwarding cost from uniform distribution in interval [1, 5]. In order to determine distribution of path durations, we computed path durations for different mobility scenarios (different mobility models, various speeds and node numbers). Then we applied Maximum Likelihood Estimation (MLE) to fit five known distributions, namely exponential, normal, lognormal, weibull and generalized pareto \cite{pappoulis:24}. The Akaike Information Criterion is used as a measure of the goodness of fit \cite{AIC:20}. The results show that exponential has the best fit among the mentioned distributions which is compatible with recent studies on distribution of path durations in mobile ad hoc networks \cite{pd1:25,pd2:26,Pathmodel:15}. We estimated parameters of exponential distribution using MLE. It should be noted that, we have estimated the parameters with a confidence level of 95\% using sufficiently large number of samples. That means, for 95\% of times the estimated parameter lies in a very small interval enclosing the real parameter.

Two scenarios are presented in this section. In the first one, 40 mobile nodes move at speeds of 1, 5, 10, 15, 20, 25, 30, 35 $m/s$. In the second case, number of mobile nodes varies between 20, 30, 40, 50, 60 and 70 while the speed is fixed at 5 $m/s$. The simulation time is set to 2000 seconds in both cases. Results are obtained by taking average among different randomly picked source-destination pairs. Between each source-destination pair, several routes with different hop numbers may exist. Obviously, for least hop routes it is more likely to have lower costs. Also, routes with greater number of hops are more likely to break \cite{Pathmodel:15}. Therefore, least hop routes will win with much higher probabilities. As a result, and without loss of generality, we only consider least-hop routes as bidders.

\begin{figure*}[!t]
\centerline{
\subfigure[For different speeds]{\label{figSavgoverpayratio}\includegraphics[width=9cm,height=6cm]{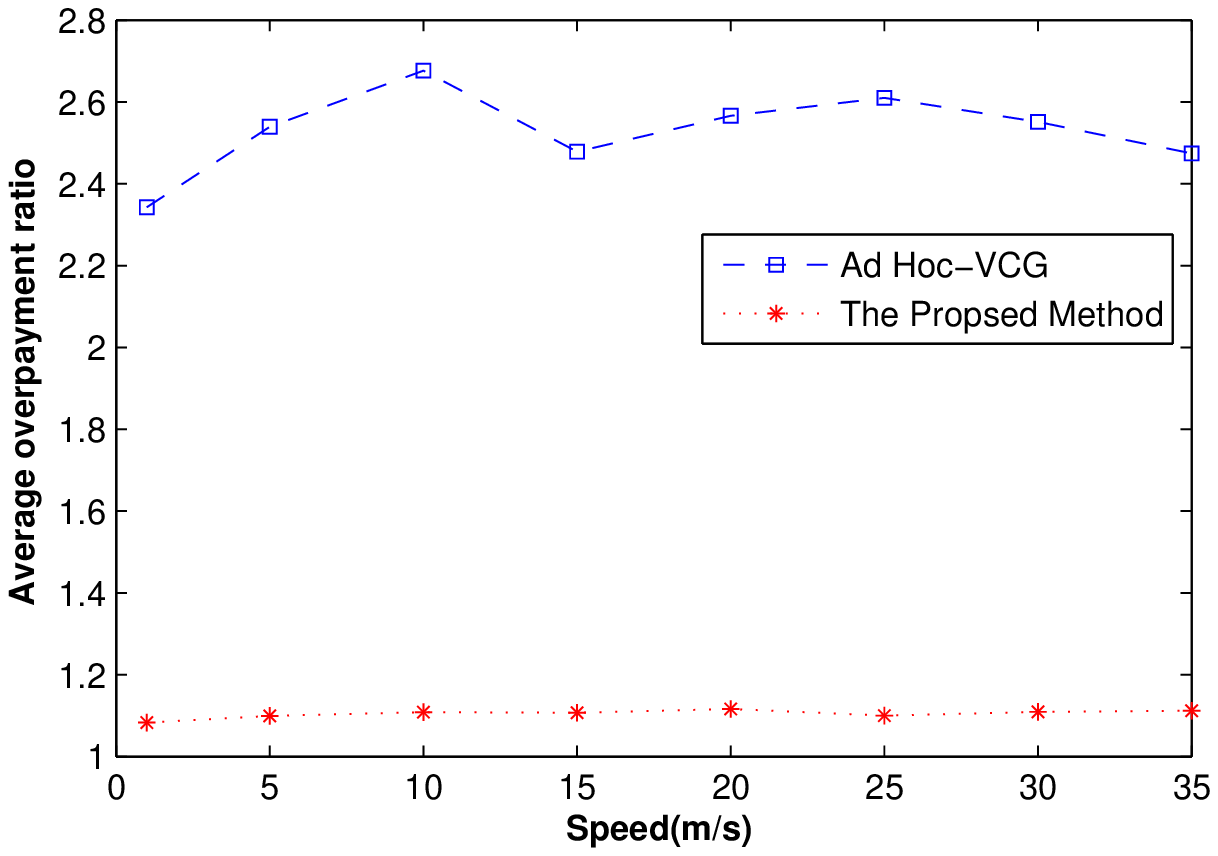}}
\hfil
\subfigure[For different number of nodes]{\label{figNavgoverpayratio}\includegraphics[width=9cm,height=6cm]{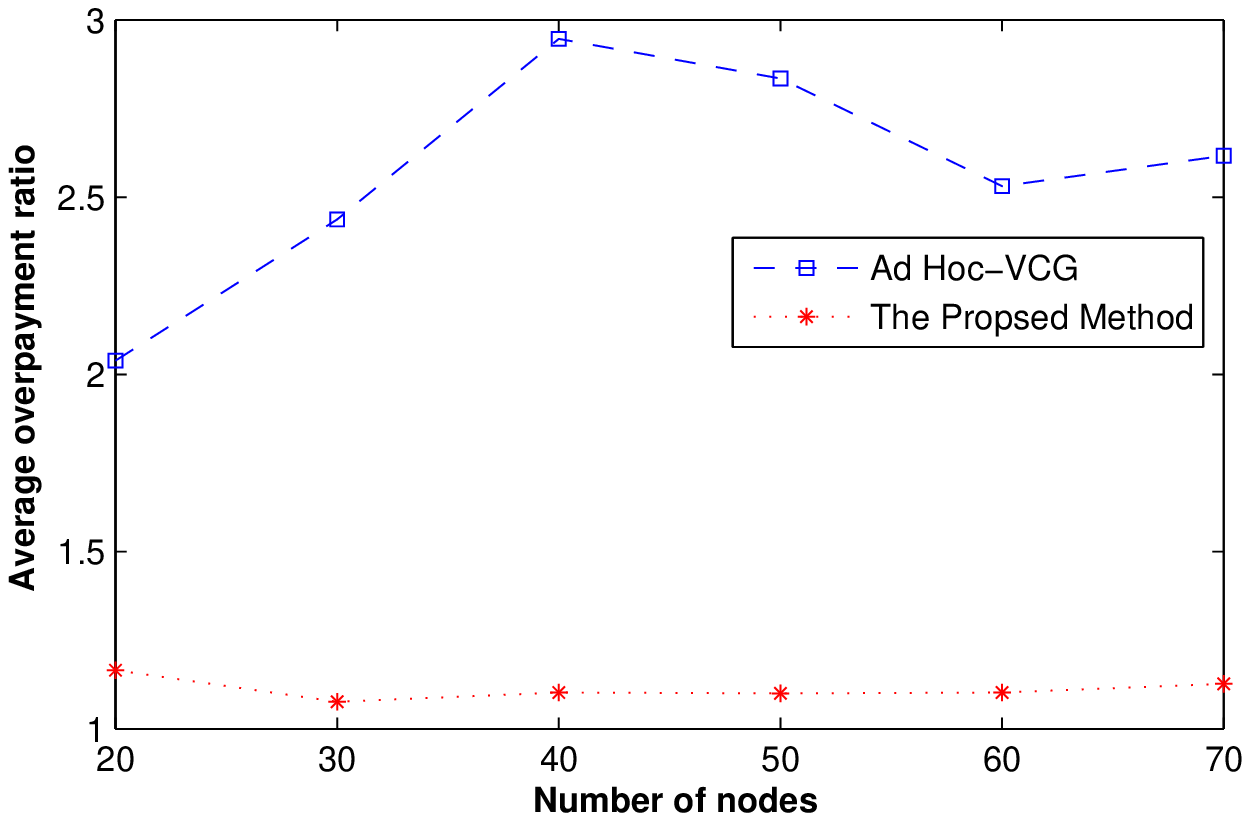}}}
\caption{Average overpayment ratio in the proposed method and Ad Hoc-VCG}
\label{figavgoverpayratio}
\end{figure*}

\begin{figure*}[!t]
\centerline{
\subfigure[For different speeds]{\label{figSmaxoverpayratio}\includegraphics[width=9cm,height=6cm]{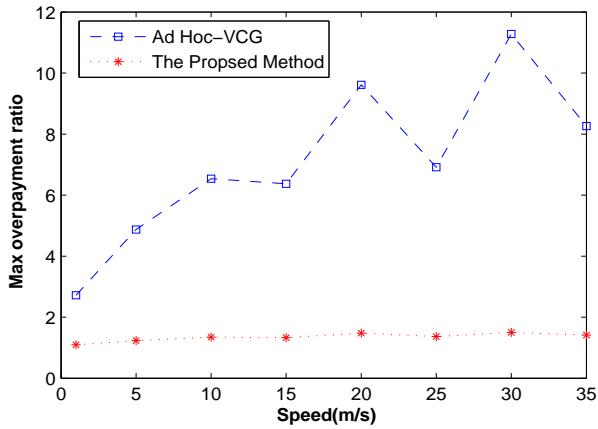}}
\hfil
\subfigure[For different number of nodes]{\label{figNmaxoverpayratio}\includegraphics[width=9cm,height=6cm]{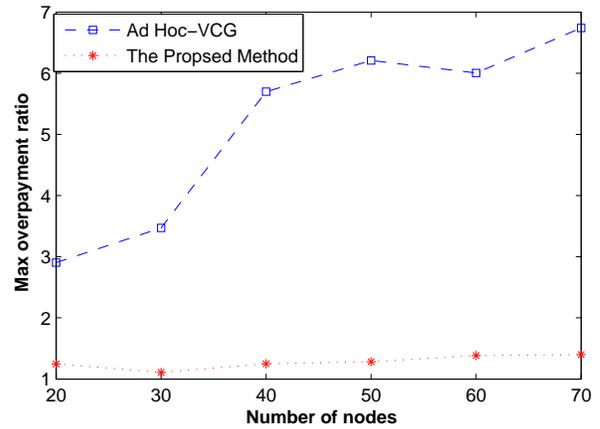}}}
\caption{Worst overpayment ratio in the proposed method and Ad Hoc-VCG}
\label{figmaxoverpayratio}
\end{figure*}

\begin{figure*}[!t]
\centerline{
\subfigure[For different speeds]{\label{figSavgpayment}\includegraphics[width=9cm,height=6cm]{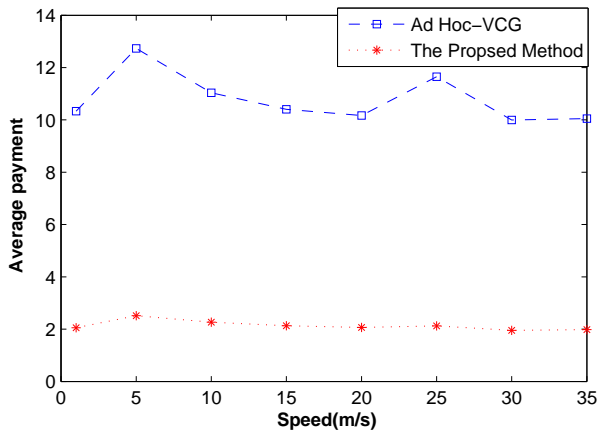}}
\hfil
\subfigure[For different number of nodes]{\label{figNavgpayment}\includegraphics[width=9cm,height=6cm]{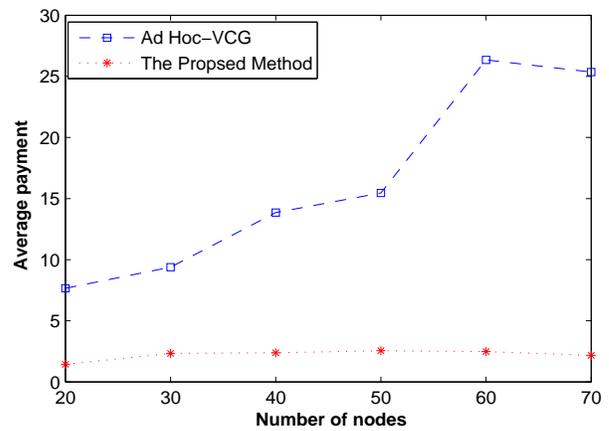}}}
\caption{Average total payment in the proposed method and Ad Hoc-VCG}
\label{figavgpayment}
\end{figure*}

In order to evaluate the proposed method, we need to define some metrics. \emph{Overpayment ratio} is obtained by dividing the total payments the auctioneer pays by the total forwarding cost. Overpayment ratio of 1 means that the auctioneer pays exactly the actual cost and the mechanism is budget balanced. Thus, it is favorable to achieve an amount close to 1 for the overpayment ratio. Fig. \ref{figSavgoverpayratio} and Fig. \ref{figNavgoverpayratio} illustrate the average overpayment ratio in our proposed auction and the Ad Hoc-VCG with changing speed of nodes and changing number of nodes respectively. It can be seen from both figures that average overpayment ratio in the proposed auction is very close to 1. On the other hand, in the Ad Hoc-VCG average overpayment ratio exceeds 2; That means the auctioneer has to pay more than twice the actual costs.

We also evaluate overpayments of our proposed auction in the worst case. For this, we define \emph{worst overpayment ratio} as the maximum overpayment ratio during the simulation time. We take average among maximum overpayment ratios of different source-destination pairs. In Fig. \ref{figSmaxoverpayratio} and Fig. \ref{figNmaxoverpayratio}, the worst overpayment ratios of the proposed auction and the Ad Hoc-VCG are compared for different speeds and different node densities respectively. As it can be seen, even in the worst case, the overpayment ratio of the proposed method is close to 1. In Ad Hoc-VCG, however, maximum of the overpayment ratios gets worse. This happens because some nodes in the selected route become pivotal in that if they are removed, the alternative route will be costly. In these situations, Ad Hoc-VCG pays enormous premiums in order to preserve the selected route.

Up to here, we evaluated the proposed auction in terms of overpayments. One may argue that overpayment corresponds to the amount of premiums, and not to the payments itself. In better words, a mechanism may select an expensive route but incur less overpayment. While, another mechanism chooses a cheaper route but pays more premiums with totally less payments. Therefore, we define another metric called \emph{Total payment} as the sum of actual costs and the premiums. This reflects the total amount the auctioneer has to pay. Fig. \ref{figSavgpayment} and Fig.\ref{figNavgpayment} compare the proposed method with the Ad Hoc-VCG in terms of the average total payment, for different speeds and different node densities respectively. As these figures show, our proposed auction pays less compared to the Ad Hoc-VCG, owing to the high premiums Ad Hoc-VCG pays. This, along with the results for overpayments, implies that our proposed mechanism is cheap and optimal in that it minimizes the total payments as well as the premiums.

\section{Conclusion}\label{conclue}

Selfish behaviors, along with mobility of nodes, make routing a difficult task in MANETs. In order to provide incentives for selfish nodes, several pricing based methods have been proposed. However, they lead to large overpayments. Moreover, most of them have ignored the mobility of nodes. Designing a mobility aware incentive compatible mechanism, which also minimizes the amount of overpayments, for routing in MANETs is a challenge that we considered in this paper. For the first time in pricing based methods, we introduced a multi-dimensional auction that allows us to take path durations, in addition to route costs, into consideration. A known mobility prediction method is used to determine path durations which reflect mobility of nodes. The proposed mechanism is also optimal in that it minimizes the amount of overpayments. Theoretical analysis is presented to support our claims. In addition, simulation results confirm the superiority of our approach over known methods in terms of payments.


%
\appendices
\section{}
\begin{lemma}
If either the $IC_d$ constraint or the $IC_u$ constraint binds, the other one is redundant.
\end{lemma}

\begin{IEEEproof}[Proof]
We show that if $IC_u$ binds, then $IC_d$ is satisfied. The reverse can be proved with an almost identical approach. $IC_u$ binds when:

\begin{displaymath}\label{lem3.1:3}
\begin{split}
p_{c,d}-p_{c+1,d}&=w(a_{c,d}|c,d)-w(a_{c+1,d}|c,d)\\
&=c(a_{c,d}-a_{c+1,d})\\
\end{split}
\end{displaymath}
Using monotonicity,
\begin{displaymath}\label{lem3.2:3}
\begin{split}
&w(a_{c,d}|c,d)-w(a_{c+1,d}|c,d)\leq w(a_{c,d}|c+1,d)-w(a_{c+1,d}|c+1,d)
\end{split}
\end{displaymath}
Therefore,
\begin{displaymath}
p_{c,d}-p_{c+1,d}\leq w(a_{c,d}|c+1,d)-w(a_{c+1,d}|c+1,d)\\
\end{displaymath}
Which is the $IC_d$ constraint.
\end{IEEEproof}

According to the $p_{c,d}$ obtained in Theorem 2:

\begin{displaymath}\label{bindic}
\begin{split}
p_{c,d}-p_{c+1,d}&=ca_{c,d}+\sum_{k=c+1}^{c_{max}}a_{k,d}-[(c+1)a_{c+1,d}+\sum_{k=c+2}^{c_{max}}a_{k,d}]\\
&=c(a_{c,d}-a_{c+1,d})\\
&=w(a_{c,d}|c,d)-w(a_{c+1,d}|c,d)
\end{split}
\end{displaymath}

Therefore, $IC_u$ binds (and similarly $IC_d$ binds). With Lemma 3, one of the $IC_d$ or $IC_u$ constraints can be eliminated. Next, we show that considering only $IC_u$ constraints yields a better result than that of $IC_d$.

Let $p_{c,d}^{up}$ and $p_{c,d}^{down}$ be payments to $(c,d)$ when only $IC_u$ and $IC_d$ are considered respectively.

\begin{displaymath}\label{ICUbetter2}
\begin{split}
p_{c,d}^{down}&=p_{c+1,d}+w(a_{c,d}|c+1,d)-w(a_{c+1,d}|c+1,d)\\
&=p_{c+1,d}+(c+1)(a_{c,d}-a_{c+1,d})
\end{split}
\end{displaymath}
\begin{displaymath}\label{ICUbetter1}
\begin{split}
p_{c,d}^{up}&=p_{c+1,d}+w(a_{c,d}|c,d)-w(a_{c+1,d}|c,d)\\
&=p_{c+1,d}+c(a_{c,d}-a_{c+1,d})\\
&p_{c,d}^{up}\leq p_{c,d}^{down}
\end{split}
\end{displaymath}

Since we seek to minimize payments, $IC_u$ yields a better payment. Therefore, we removed $IC_d$ constraints from the graph.

\section{}
\begin{lemma}
If allocation rule $a$ is monotonic then $a$ is incentive compatible.
\end{lemma}
\begin{IEEEproof}[Proof]
According to the $p_{c,d}$ obtained in Theorem 2:
\begin{displaymath}
\begin{split}
p_{c,d}-p_{c+1,d}&=ca_{c,d}+\sum_{k=c+1}^{c_{max}}a_{k,d}-[(c+1)a_{c+1,d}+\sum_{k=c+2}^{c_{max}}a_{k,d}]\\
&=c(a_{c,d}-a_{c+1,d})\\
&=w(a_{c,d}|c,d)-w(a_{c+1,d}|c,d)
\end{split}
\end{displaymath}
Therefore, $IC_u$ binds. Again, by using $p_{c,d}$ obtained in Theorem 2:

\begin{displaymath}\label{pathic}
p_{c,d}-p_{c,d-1}=ca_{c,d}+\sum_{k=c+1}^{c_{max}}a_{k,d}-[ca_{c,d-1}+\sum_{k=c+1}^{c_{max}}a_{k,d-1}]
\end{displaymath}

According to monotonicity, $\sum_{k=c+1}^{c_{max}}a_{k,d}-\sum_{k=c+1}^{c_{max}}a_{k,d-1}\geq 0$. Therefore,\\
\begin{displaymath}
\begin{split}
&p_{c,d}-p_{c,d-1}\geq ca_{c,d}-ca_{c,d-1}\\
&p_{c,d}-w(a_{c,d}|c,d)\geq p_{c,d-1}-w(a_{c,d-1}|c,d)
\end{split}
\end{displaymath}
By Lemma 3 and Theorem 1, the allocation rule $a$ is incentive compatible.
\end{IEEEproof}
\begin{lemma}
Allocation rule $a$ that is incentive compatible must be monotonic.
\end{lemma}
\begin{IEEEproof}
First, assume that $c' \geq c$ but $a_{c',d} > a_{c,d}$. IC constraint $(c',d) \rightarrow (c,d)$ implies that
\begin{displaymath}
\begin{split}
&p_{c',d}-p_{c,d}\geq w(a_{c',d}|c',d)-w(a_{c,d}|c',d)\\
&w(a_{c',d}|c',d)-w(a_{c,d}|c',d)\geq w(a_{c',d}|c,d)-w(a_{c,d}|c,d)\\
&p_{c',d}-p_{c,d}\geq w(a_{c',d}|c,d)-w(a_{c,d}|c,d)
\end{split}
\end{displaymath}
That means $p_{c',d}-w(a_{c',d}|c,d)\geq p_{c,d}-w(a_{c,d}|c,d)$ violates IC.

Next, assume that $d'\leq d$ but $a_{c',d'}>a_{c',d}$. IC constraint $(c',d) \rightarrow (c',d')$ implies that
\begin{displaymath}
\begin{split}
&p_{c',d}-p_{c',d'}\geq w(a_{c',d}|c',d)-w(a_{c',d'}|c',d)\\
&w(a_{c',d}|c',d)-w(a_{c',d'}|c',d) = w(a_{c',d}|c',d')-w(a_{c',d'}|c',d')\\
&p_{c',d}-p_{c',d'}\geq w(a_{c',d}|c',d')-w(a_{c',d'}|c',d')
\end{split}
\end{displaymath}
That means $p_{c',d}-w(a_{c',d}|c',d')\geq p_{c',d'}-w(a_{c',d'}|c',d')$ violates IC.
\end{IEEEproof}
\begin{theorem}
Allocation rule $a$ is monotonic if and only if $a$ is incentive compatible.
\end{theorem}
\begin{IEEEproof}
With Lemma 4 and Lemma 5 the theorem is proven.
\end{IEEEproof}



\ifCLASSOPTIONcaptionsoff
  \newpage
\fi



%
\bibliographystyle{IEEEtran}
\bibliography{IEEEabrv}



%








\end{document}